\documentclass[lettersize,journal]{IEEEtran}
\usepackage{amsmath,amsfonts}
\usepackage{algorithmic}
\usepackage{algorithm}
\usepackage{array}
\usepackage[caption=false,font=normalsize,labelfont=sf,textfont=sf]{subfig}
\usepackage{textcomp}
\usepackage{stfloats}
\usepackage{url}
\usepackage{verbatim}
\usepackage{graphicx}
\usepackage{cite}
\usepackage{multirow}
\begin{document}

\title{On-Device Super Resolution Imaging Using Low-Cost SPAD Array and Embedded Lightweight Deep Learning}

\author{Zhenya~Zang$^\dagger$,
        Xingda~Li$^\dagger$,
        and~David~Day~Uei~Li%
\thanks{Z. Zang, X. Li, and D. D. U. Li are with the Department of Biomedical Engineering, University of Strathclyde, 16 Richmond Street, Glasgow, G1 1XQ, United Kingdom. The authors acknowledge the support from Datalab, Photon Force, Xilinx, the EPSRC (EP/T00097X/1), the Quantum Technology Hub in Quantum Imaging (QuantiC), and Innovate UK HYDRI (10005391).}
\thanks{$\dagger$ Authors contributed equally to this work.}
\thanks{Corresponding author: David Day Uei Li (david.li@strath.ac.uk).}}
\markboth{}%
{Zang \MakeLowercase{\textit{et al.}}: On-Device Super Resolution Imaging Using Low-Cost SPAD Array and Embedded Lightweight Deep Learning}

\maketitle

\begin{abstract}
This work presents a lightweight super-resolution (LiteSR) neural network for depth and intensity images acquired from a consumer-grade single-photon avalanche diode (SPAD) array with a 48$\times$32 spatial resolution. The proposed framework reconstructs high-resolution (HR) images of size 256$\times$256. Both synthetic and real datasets are used for performance evaluation. Extensive quantitative metrics demonstrate high reconstruction fidelity on synthetic datasets, while experiments on real indoor and outdoor measurements further confirm the robustness of the proposed approach. Moreover, the SPAD sensor is interfaced with an Arduino UNO Q microcontroller, which receives low-resolution (LR) depth and intensity images and feeds them into a compressed, pre-trained deep learning (DL) model, enabling real-time SR video streaming. In addition to the 256$\times$256 setting, a range of target HR resolutions is evaluated to determine the maximum achievable upscaling resolution (512$\times$512) with LiteSR, including scenarios with noise-corrupted LR inputs. The proposed LiteSR-embedded system co-design provides a scalable, cost-effective solution to enhance the spatial resolution of current consumer-grade SPAD arrays to meet HR imaging requirements.
\end{abstract}


\section{Introduction}
\IEEEPARstart{S}{ingle}-photon avalanche diode (SPAD) arrays with time-correlated single-photon counting (TCSPC) modules are prevalent in machine vision for autonomous systems \cite{rapp2020advances} and life sciences \cite{williams2021full,li2010real}, due to their high single-photon sensitivity and time-of-flight (TOF) measurement capabilities. Benefiting from advances in semiconductor technology, the spatial resolution of consumer-grade SPAD arrays has consistently improved, enabling compact sensors with increasingly larger array sizes: from 8 $\times$ 8 \cite{st_vl53l9cx} to 48 $\times$ 32 \cite{ams_tmf8829} on the market and 160 $\times$ 120 \cite{kappel2024monolithic} in research. Low-cost 8 $\times$ 8 arrays have also been used for fluorescence lifetime imaging (FLIM) with a scanning mechanism \cite{wilson2025high}, achieving spatial resolution comparable to that of commercial FLIM systems, though with reduced temporal resolution ($800\,\mathrm{ps} \ \text{vs.}\ <10\,\mathrm{ps}$).

Beyond time-resolved imaging, the spatial resolution of SPAD arrays fundamentally limits their applicability in depth sensing and scene understanding tasks, motivating computational SR approaches. Deep learning (DL)-based computational SR methods have been widely adopted for SR imaging with low-cost SPAD sensors, as they improve resolution without requiring a physically larger pixel array. Most existing SR methods for consumer-grade SPAD arrays fuse low-resolution (LR) depth maps from 8$\times$8 SPAD arrays with high-resolution (HR) guidance images from indirect time-of-flight (iToF) cameras \cite{ruget2022pixels2pose} or color images from RGB cameras \cite{wilson2025high,ding2025cfpnet,xiang2025depthor,zang2025deep} by learning to capture HR spatial information. Depth SR and completion are also facilitated by 8$\times$8 SPAD arrays \cite{ding2025cfpnet,xiang2025depthor}. Some non-fusion strategies demonstrate the feasibility of using LR SPAD alone, using images \cite{zang2025deep}, histograms \cite{li2025facial}, or temporal representations of histograms \cite{wilson2025superresolution} from the 8 $\times$ 8 SPAD sensors to generate HR images.

Despite the success of non-fusion SR imaging, upscaling an 8$\times$8 image to an HR representation is extremely challenging due to the bottleneck imposed by the ultra-low spatial sampling. An 8$\times$8 input provides insufficient spatial context to reliably infer fine-scale structures, edges, and textures present in the target HR image. Therefore, the learning task becomes highly ill-posed, with multiple plausible HR reconstructions corresponding to the same LR input. The recently introduced 48 $\times$ 32 SPAD sensor (TMF8829), with its increased pixel count, significantly alleviates this bottleneck. Unlike prior SPAD SR works that target depth alone, the TMF8829 simultaneously outputs depth and intensity maps, enabling joint SR of both modalities. Meanwhile, modern embedded platforms with sufficient on-board memory and computational resources enable on-the-fly SR processing of sensor outputs from the SPAD array and provide customizable firmware and sufficient interfaces for integration with the sensor. In this work, we propose a lightweight SR network tailored to the 48×32 TMF8829 sensor, designed for efficient inference on embedded devices via low-bit quantization. The contribution of this work is threefold:

We leverage the higher spatial resolution of a TMF8829 SPAD array to propose LiteSR, a lightweight dual-branch DL model for non-fusion depth and intensity SR. Unlike fusion-based approaches, LiteSR processes internal LR depth and intensity (photon count) inputs through dedicated branches, each tailored to the characteristics of its respective modality, without relying on HR guidance from external cameras, and generates HR outputs at 256$\times$256 and up to 512$\times$512 pixels. Quantization-aware training (QAT) is applied to produce a compact INT8 model, demonstrating both accuracy retention and model size reduction for downstream deployment on resource-constrained embedded hardware.

Extensive evaluations are conducted on both synthetic and real data. Multiple metrics for upscaled intensity and depth images are used to demonstrate accuracy and robustness. Ablation studies, including the removal of specific modules and maximum upscaling stress tests, are conducted to validate design choices and demonstrate LiteSR's flexibility.

LiteSR is implemented on an Arduino UNO Q with 2 GB of RAM, paired with a TMF8829 sensor, and uses customized firmware to handle sensor configuration, data flow, and on-device LiteSR inference. SR depth and intensity images can be streamed from the system in real time. Both stationary and dynamic scenes are measured.

\section{Data Generation and Acquisition}
Synthetic training datasets are generated using NYUv2 \cite{silberman2012indoor}. Details of data generation and non-illuminated dead-zone compensation are described in our previous study \cite{zang2025deep}, with the key differences that the LR input resolution is 32$\times$48 rather than 8$\times$8 and that 10 more scenes are included. In total, 3,121 pairs of LR and HR intensity and depth images of indoor scenes are generated and split into training and validation sets (2,497 and 624 pairs, respectively). Pixel-to-pixel response non-uniformity calibration maps for depth and intensity, measured using a white flat target, are applied in the data loader during training to simulate realistic sensor captures. Independent test datasets are generated from the Middlebury Stereo Datasets \cite{scharstein2007learning}. Realistic datasets are collected using a TMF8829 SPAD sensor configured with a 32$\times$48 spatial resolution, 64 time bins, and an 800 ps temporal bin width, operating in high-accuracy mode with a maximum measurement range of 1.4 m. The Arduino board interfaces with the TMF8829 via SPI using a modified C++-based Arduino TMF8829 library \cite{ams2025tmf8829}.

\begin{figure*}
\centering
\includegraphics[width=2\columnwidth]{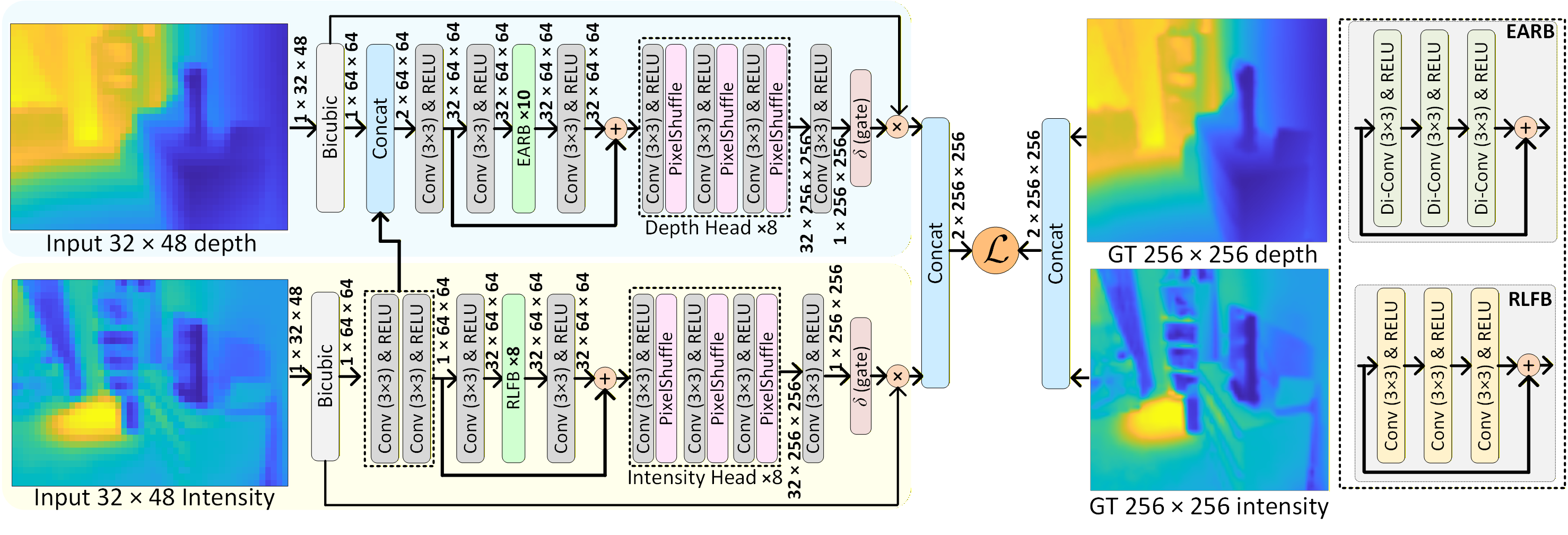}
\caption{LiteSR architecture, composed of intensity-guided depth and intensity branches. The numbers of channels for the EARB, RLFB, Depth Head, and Intensity Head are denoted by CH\_E, CH\_R, CH\_DH, and CH\_IH.}
\label{fig:litesr-architecture}
\end{figure*}

\section{Deep Learning Architecture}
Given that LiteSR is intended for deployment on an Arduino-based embedded device, it is constructed using plain and dilated convolutional neural network (CNN) modules and skip connections, without complex backbones, to ensure simplicity and compatibility with resource-constrained embedded hardware. As shown in Fig.~\ref{fig:litesr-architecture}, the proposed model adopts a dual-branch architecture, where each branch reconstructs 256 $\times$ 256 HR depth and intensity images from 32 $\times$ 48 LR depth and intensity inputs, respectively. The model follows an interpolation-based residual learning framework, where a residual connection links a bicubic interpolation baseline at the input to the final network output. This design provides reliable low-frequency information while enabling learning-based modules to focus on recovering high-frequency details. The reconstruction is generally formulated as
\begin{equation}
y = y_{\text{base}} + \delta \cdot r,
\label{eq:reconstruction}
\end{equation}
where $y_{\text{base}}$ denotes the bicubic interpolation baseline, $r$ represents the learned residual features predicted by the CNN branches, and $\delta$ is a learnable sigmoid gate that adaptively regulates the contribution of the residuals. This gating mechanism stabilizes optimization by initially favoring interpolation and progressively increasing neural corrections only when beneficial. The depth and intensity reconstruction branches adopt different architectural designs to reflect their distinct feature-extraction characteristics. The depth branch emphasizes geometric structure and depth discontinuities, while the intensity branch focuses on refining local textural detail. To exploit the structural correlation between modalities, the intensity signal is used to guide depth reconstruction, as shown in the dashed box in Fig.~\ref{fig:litesr-architecture}. Specifically, a lightweight guidance module extracts an edge-aware mask from the LR intensity input, concatenates it with the LR depth features, and feeds the result into the depth residual branch. This strategy improves edge preservation in depth reconstruction without coupling the two prediction heads, thereby preventing cross-modal feature interference arising from the fundamentally different spatial-frequency distributions and measurement modalities of depth and intensity signals.

The depth branch employs feature fusion followed by a stack of Edge-Aware Residual Blocks (EARB) with dilation convolutions, enabling effective long-range geometric context modeling. PixelShuffle modules \cite{shi2016realtime} are used for learnable sub-pixel upsampling in both branches, enabling the reconstruction of high-frequency detail while avoiding interpolation-induced smoothing. In contrast, the intensity branch replaces the edge-aware blocks with Residual Local Feature Blocks (RLFBs), which are better suited for extracting and refining local spatial textures. The total loss is a weighted sum of the depth branch loss and the intensity branch loss:
\begin{equation}
L = w_d \cdot L_d + w_i \cdot L_i,
\label{eq:total-loss}
\end{equation}
where $w_d = 0.05$ and $w_i = 0.1$ are the branch weights, fine-tuned to the optimal values. Depth branch loss is defined as
\begin{equation}
L_d = L_{\text{charb}} + w_{\text{SSIM}} \cdot (1 - L_{\text{SSIM}})
+ w_{\text{edge}} \cdot L_{\text{edge}} + w_{\text{TV}} \cdot L_{\text{TV}},
\label{eq:depth-loss}
\end{equation}
where Charbonnier loss \cite{charbonnier1994deterministic}
\begin{equation}
L_{\text{charb}} = \frac{1}{N} \sum \sqrt{(\hat{y} - y)^2 + \varepsilon^2},
\label{eq:charbonnier}
\end{equation}
calculates smooth L1 loss between predicted depth $\hat{y}$ and ground-truth (GT) depth image $y$, with $\varepsilon = 0.001$ for numerical stability of gradient computation. The Structural Similarity Index (SSIM) loss $L_{\text{SSIM}}$ with $w_{\text{SSIM}} = 0.2$ measures structural differences. The Sobel edge loss with $w_{\text{edge}} = 0.2$ penalizes differences in edge structure between the predicted and GT depth maps, not only preserving high-frequency edge fidelity but also compensating for the effect of pixel-wise artifacts from $L_{\text{charb}}$. The total variation (TV) loss \cite{rudin1992nonlinear} $L_{\text{TV}}$ with $w_{\text{TV}} = 10^{-5}$ encourages spatial smoothness in the predicted depth map. Similarly, the intensity branch loss is defined as
\begin{equation}
L_i = L_{\text{charb}} + w_{\text{MS-SSIM}} \cdot \left(1 - L_{\text{MS-SSIM}}\right)
+ w_{\text{grad}} \cdot L_{\text{grad}},
\label{eq:intensity-loss}
\end{equation}
which also includes $L_{\text{charb}}$, but with Multi-Scale SSIM (MS-SSIM) loss $L_{\text{MS-SSIM}}$ \cite{wang2003multiscale} and $w_{\text{MS-SSIM}} = 0.15$ because intensity images contain richer textural details and greater perceptual sharpness than depth images. The Sobel gradient loss $L_{\text{grad}}$ with $w_{\text{grad}} = 0.02$ preserves edge and textural-gradient fidelity.

To validate the feasibility of implementing LiteSR on embedded hardware, INT8 QAT is applied to evaluate both model size and performance under low-precision constraints. The model is implemented in PyTorch and trained on an NVIDIA GeForce RTX 5070 Ti GPU, taking about 1 hour, with a batch size of 16. LR and corresponding HR images are normalized using the statistical mean and standard deviation computed from the training datasets. Data augmentation is applied on horizontal and vertical flips. AdamW \cite{loshchilov2019decoupled} is the optimizer with an initial learning rate of $2\times10^{-4}$, weight decay of $10^{-3}$, and optimizer momentum coefficients $\beta=(0.9,0.99)$. Training is performed for 200 epochs. A cosine learning-rate decay method is used, with a minimum ratio of $5\times10^{-4}$. Gradient norms are clipped to 1.0 to stabilize training.

\section{Offline Evaluation on Synthetic Datasets}
\begin{figure*}
\centering
\includegraphics[width=2\columnwidth]{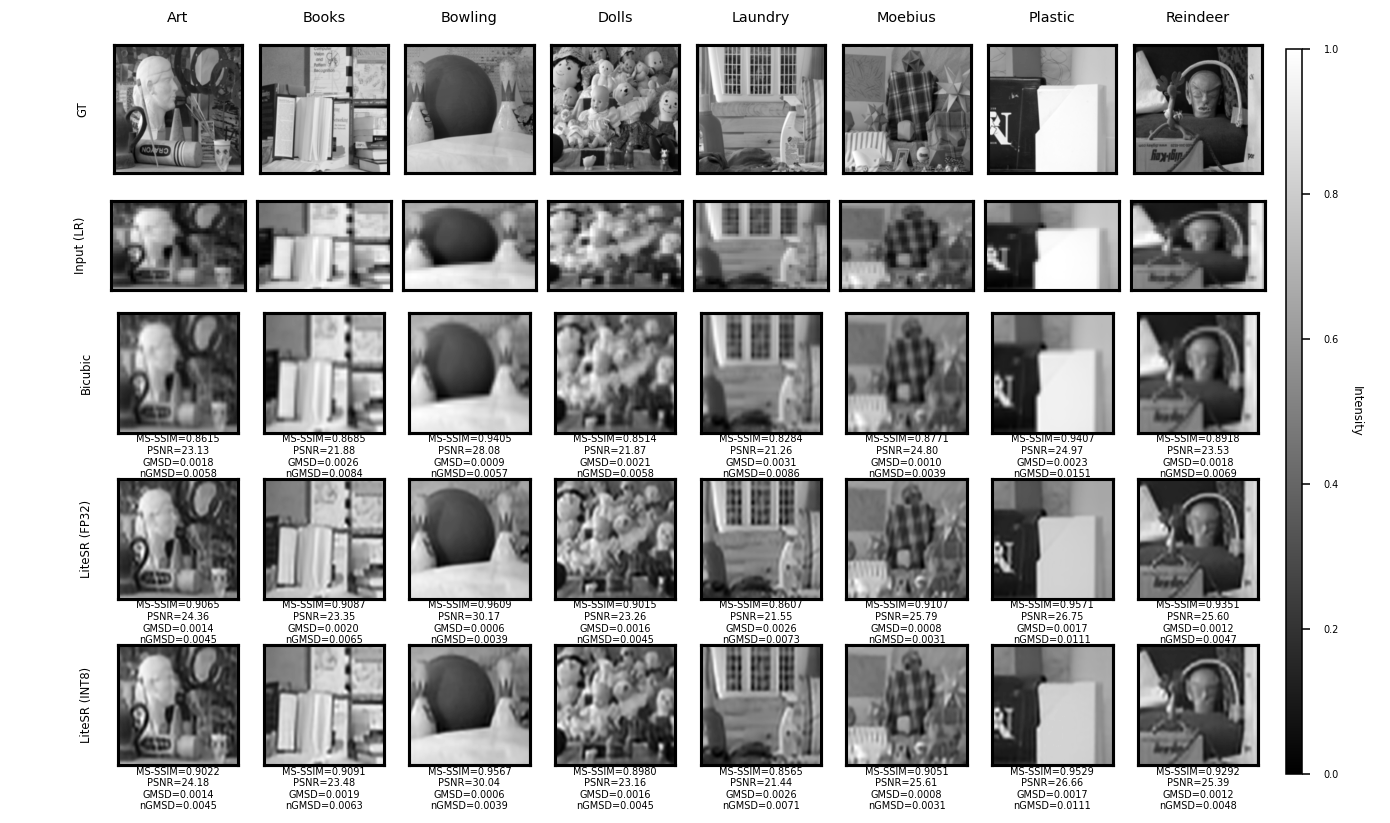}
\caption{Comparison of super-resolution results on intensity test images for $4\times$ upscaling. From left to right: GT, LR, bicubic interpolation, LiteSR (FP32), and LiteSR (INT8). Quantitative metrics (MS-SSIM, PSNR, and GMSD) are reported for each reconstructed image. The input LR image has dimensions 32x48, resulting in different aspect ratios for the output HR images.}
\label{fig:synthetic-intensity-4x}
\end{figure*}

\begin{figure*}
\centering
\includegraphics[width=2\columnwidth]{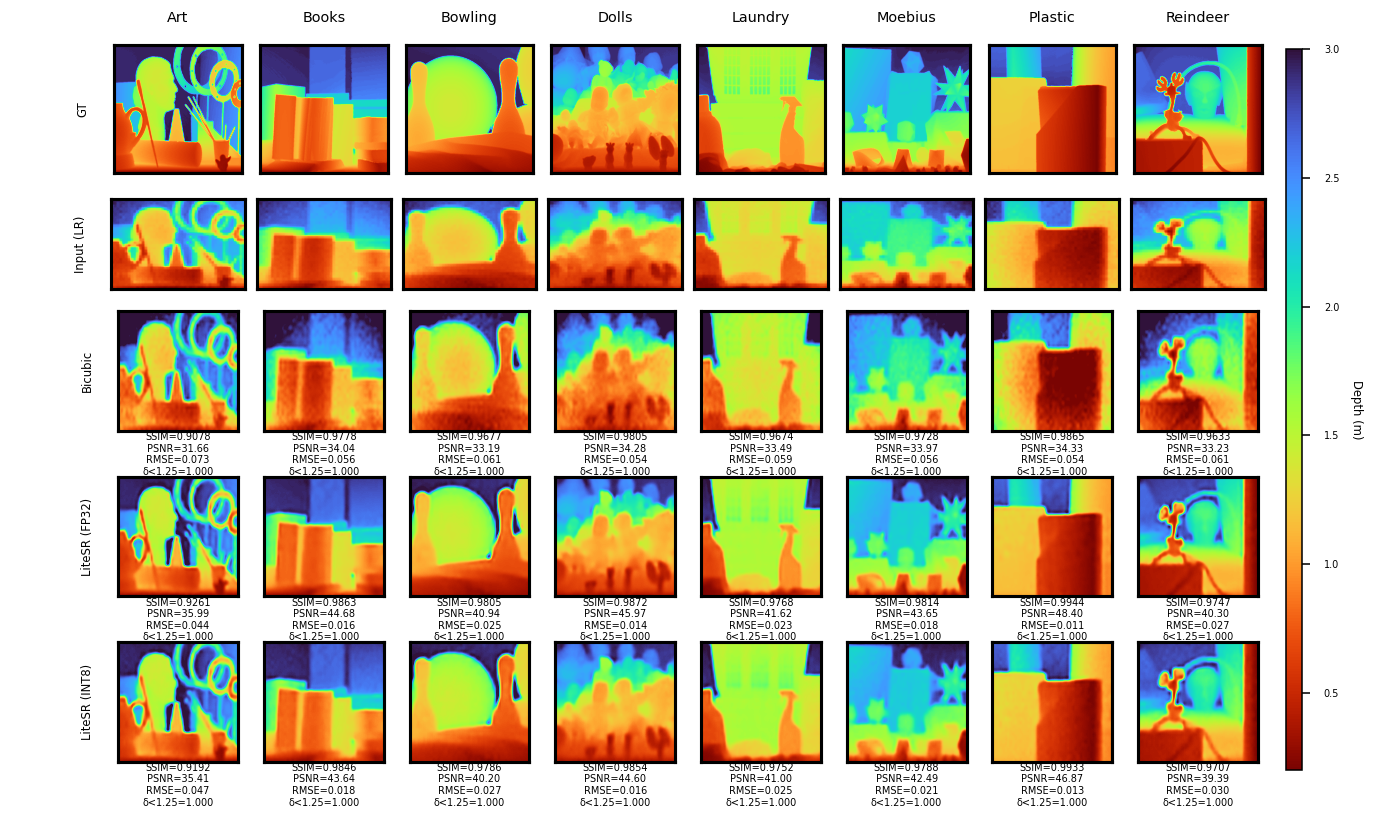}
\caption{Comparison of super-resolution results on depth test images for $4\times$ upscaling. From left to right: GT, LR, bicubic interpolation, LiteSR (FP32), and LiteSR (INT8). The Input LR image has dimensions 32x48, resulting in different aspect ratios for the output HR images.}
\label{fig:synthetic-depth-4x}
\end{figure*}

Multiple metrics are used to evaluate the upscaled depth and intensity images, as shown in Figs.~\ref{fig:synthetic-intensity-4x} and \ref{fig:synthetic-depth-4x}. The evaluation includes ground-truth (GT) images, LR input images, bicubic-interpolated images, outputs from the trained floating-point (FP32) model, and outputs from the INT8 model. For intensity-image evaluation, MS-SSIM is adopted as it better captures perceptual image quality across multiple spatial scales, making it more sensitive to texture consistency and contrast variations than single-scale SSIM. The peak signal-to-noise ratio (PSNR) is reported as a complementary metric for quantifying pixel-wise reconstruction fidelity. Gradient Magnitude Similarity Deviation (GMSD) \cite{xue2013gradient} is used to evaluate edge and structural fidelity by measuring the consistency of gradient-magnitude similarity between reconstructed and reference images. To enable fair comparison across images with varying dynamic ranges, normalized GMSD (nGMSD) is also reported, providing a scale-invariant measure of gradient-based structural distortion. For depth-image evaluation, SSIM is used to assess structural consistency, while a metric $\delta < 1.25$ is reported to evaluate scale-invariant depth accuracy.

In the eight test scenes, LiteSR outperforms bicubic interpolation for both depth and intensity reconstruction, producing sharper, more coherent depth discontinuities, as evidenced by higher depth SSIM and PSNR, lower RMSE, and consistently strong $\delta < 1.25$ scores. For intensity images, LiteSR retrieves finer local textures and contrast, leading to higher MS-SSIM and PSNR and lower GMSD and nGMSD, confirming improved perceptual and structural fidelity. The INT8-quantized model closely matches the FP32 model in both visual quality and quantitative metrics, with only marginal performance degradation while substantially reducing model size and computational cost. In terms of model size, FP32 requires 1.36 MB and 4.086 GFLOPs, whereas INT8 reduces these to 363.80 KB and 1.945 GIOPs, demonstrating the feasibility of implementation on resource-constrained embedded hardware.

\section{Offline Evaluation of Real Datasets}
\begin{figure}[!t]
\centering
\includegraphics[width=\columnwidth]{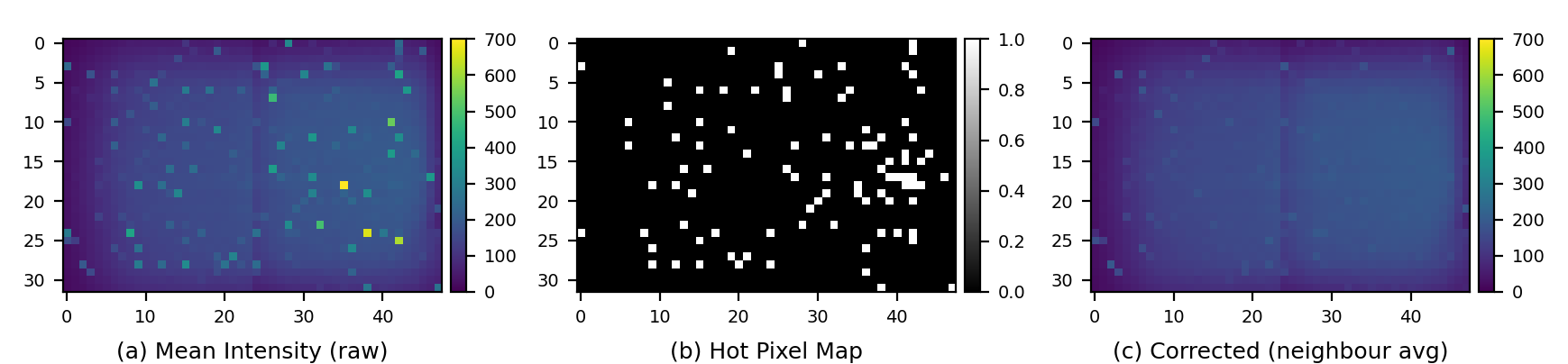}
\caption{Hot pixel detection and compensation in photon counting mode using a white flat target. (a) Raw mean intensity map showing spatially distributed hot pixels with anomalously high DCRs. (b) Binary hot-pixel mask generated by thresholding at 250 photon counts. (c) Corrected intensity image obtained by replacing hot pixels with the local $3\times3$ neighborhood average.}
\label{fig:hot-pixel-compensation}
\end{figure}

Hot pixels in the SPAD array affect the visualization of intensity mode due to the high dark count rate (DCR). Figure~\ref{fig:hot-pixel-compensation} illustrates a three-stage hot pixel compensation pipeline using a uniform white flat-field scene, where pixels whose DCR exceeds a fixed threshold are flagged and replaced by the median of their immediate 3$\times$3 neighbors. The hot pixel map is computed offline once during calibration and stored on-device, and applied to each LR intensity image prior to LiteSR inference. The smallest model (CH\_E=10, CH\_R=5, CH\_DH=6, and CH\_IH=4) is used for offline evaluation to assess worst-case reconstruction quality under the most constrained inference budget. Since no HR GT images are available, Laplacian variance (LV) \cite{pertuz2013analysis} is adopted to evaluate HR intensity reconstruction, as it captures high-frequency components such as edges and fine textural details. A gradient-based, reference-free metric is also adopted to measure depth variance within homogeneous regions, where lower values indicate smoother regions and greater reconstruction stability.

\begin{figure}[!t]
\centering
\includegraphics[width=\columnwidth]{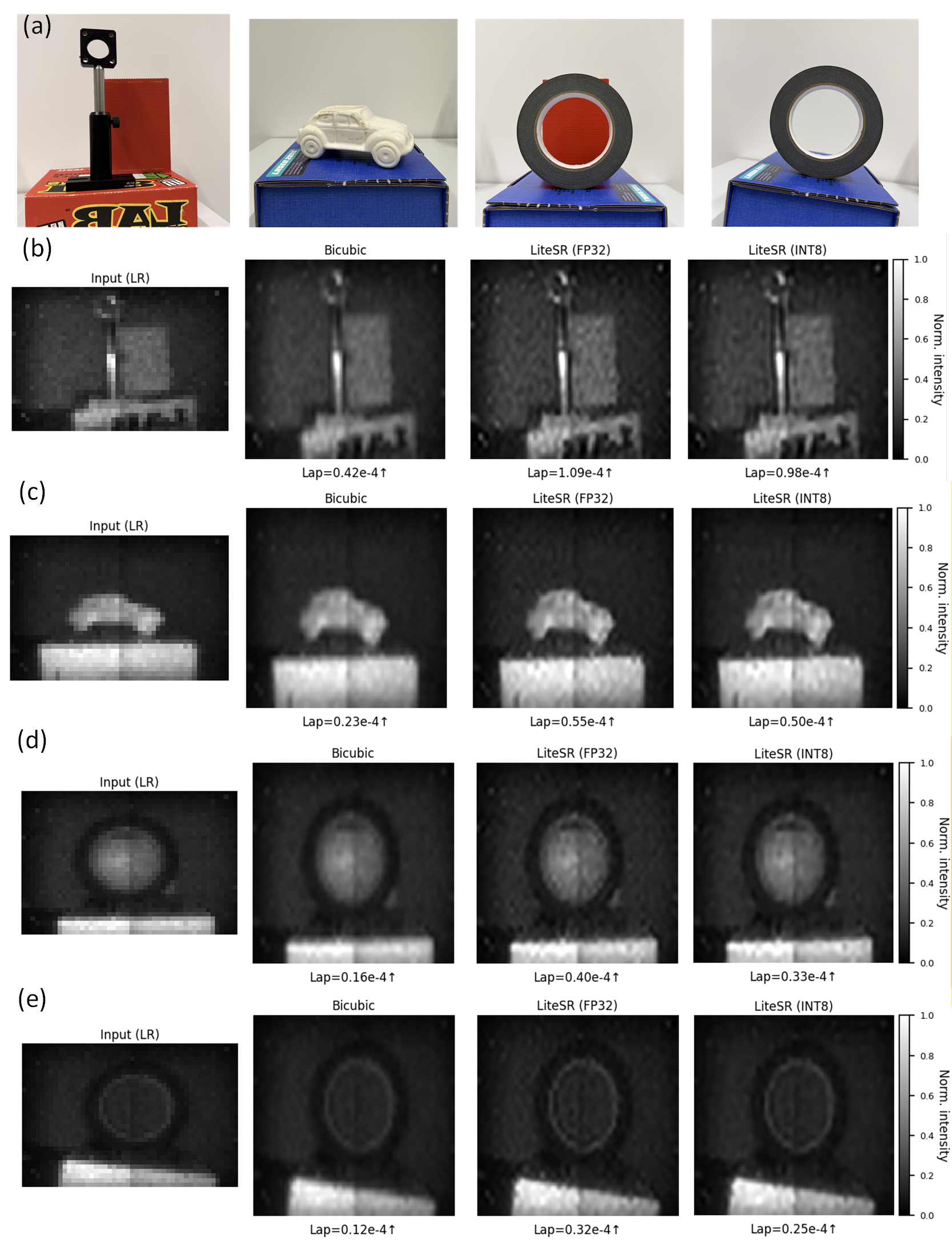}
\caption{SR results on real-world intensity captures. RGB views of the scenes are shown together with bicubic, LiteSR (FP32), and LiteSR (INT8) reconstructions and their LV scores.}
\label{fig:real-intensity-results}
\end{figure}

\begin{figure}[!t]
\centering
\includegraphics[width=\columnwidth]{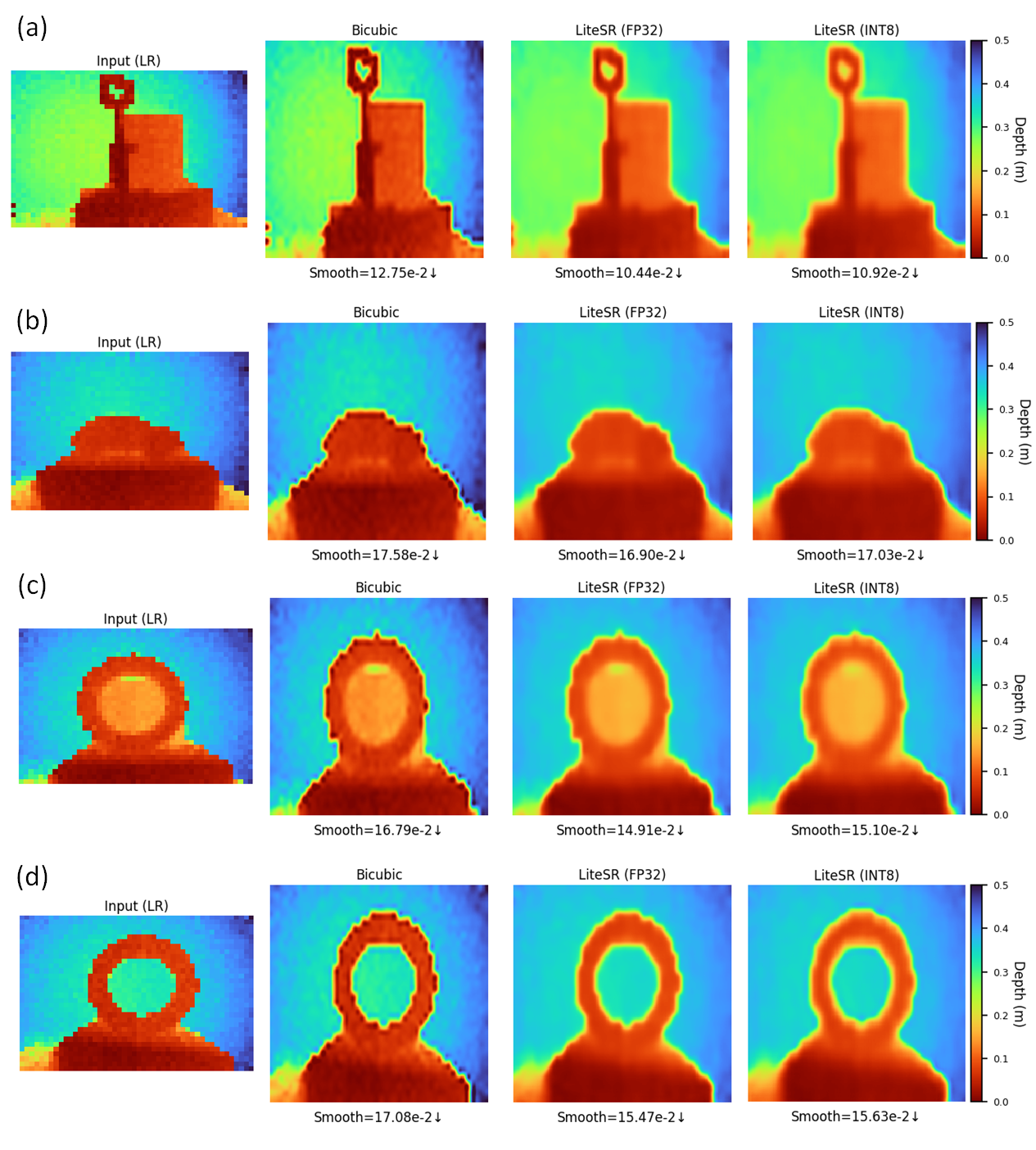}
\caption{SR results on real-world depth maps. Bicubic, LiteSR (FP32), and LiteSR (INT8) outputs are compared using depth smoothness scores.}
\label{fig:real-depth-results}
\end{figure}

The inferred HR intensity and depth images, along with quantitative comparisons, are shown in Figs.~\ref{fig:real-intensity-results} and \ref{fig:real-depth-results}, respectively. LiteSR (FP32) consistently recovers sharper edges and finer details than bicubic interpolation, as indicated by the higher LV values across all scenes, while INT8 quantization preserves most of the high-frequency content with only a slight reduction in LV. It is worth noting that, as shown in Fig.~\ref{fig:real-depth-results}, bicubic interpolation preserves the original sensor-dependent bias and contrast, whereas the SR outputs deviate from these characteristics. This is because bicubic directly upsamples the uncalibrated LR measurement. In contrast, as shown in Fig.~\ref{fig:synthetic-depth-4x}, our model is trained with pixel-wise response non-uniformity corruption and is expected to remove these effects during inference. Therefore, its output is not expected to match the LR input’s contrast exactly but rather to reconstruct a more spatially coherent depth map.

\section{Embedded System Integration and Online Processing}
\begin{figure}[!t]
\centering
\includegraphics[width=\columnwidth]{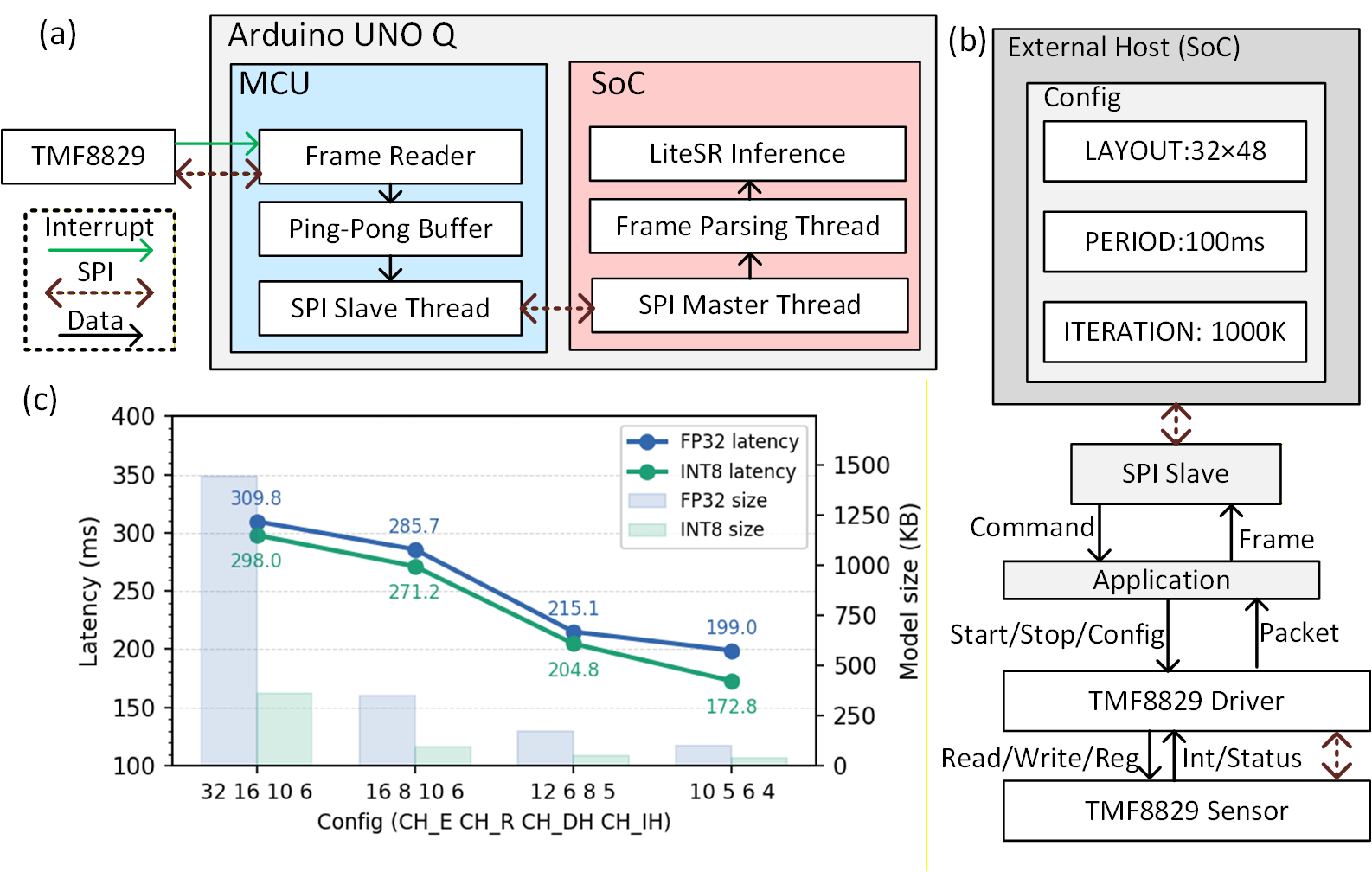}
\caption{System overview of the TMF8829-based depth sensing pipeline. (a) System architecture. (b) Software stack for configuration, acquisition, and sensor control. (c) Inference latency and model size for FP32 and INT8 LiteSR models across four channel configurations.}
\label{fig:embedded-system-overview}
\end{figure}

The hardware implementation of the embedded inference system is shown in Fig.~\ref{fig:embedded-system-overview}(a), together with the customized firmware architecture. A TMF8829 sensor is connected over SPI to an Arduino UNO Q that features an STM32U585 MCU and a Qualcomm QRB2210 SoC. On the MCU side, a frame reader captures incoming sensor data and places it into a ping-pong buffer, while an SPI slave thread makes those buffered frames available to the SoC. Sensor data readiness is indicated by an interrupt signal from the TMF8829. On the SoC side, an SPI master thread fetches frames from the MCU. A frame-parsing thread interprets the data to reconstruct the distance and intensity frames, and the LiteSR inference block implemented with on-board PyTorch performs online SR processing on the parsed frames. Inter-processor communication is handled through the Remote Procedure Call (RPC) layer, which is well-suited for control and general messaging but not for continuous transfer of dense sensor frames. Therefore, a custom high-throughput SPI transport is implemented to stream TMF8829 frame data.

Fig.~\ref{fig:embedded-system-overview}(b) illustrates a high-level abstraction of the TMF8829 sensor configuration and data acquisition flow within the system. The host provides measurement configuration parameters, including layout, period, and iteration count, as well as the controlling commands, through the custom SPI interface. The application layer translates these commands into start, stop, and configuration requests for the TMF8829 driver, which communicates with the sensor through SPI to register access, status monitoring, and interrupt handling. During ranging, the driver reads raw measurement data from the sensor’s FIFO, while the application layer formats the data into a framed packet with a header and fills the SPI slave buffer for host retrieval. In this implementation, the sensor is configured with a 32×48 layout, an iteration setting of 1000k, and a measurement period of 100 ms, which corresponds to a nominal frame rate of 10 Hz. To evaluate the hardware performance, Fig.~\ref{fig:embedded-system-overview}(c) compares latency and model size across different model configurations at FP32 and INT8 precision, showing the trade-off. With a smaller channel configuration and quantization, the model size and computational cost decrease, yielding the best model size and latency of 36.77 KB and 172.8 ms (${\sim}5.8$ fps), respectively. However, due to the visualization overhead, the observed runtime frame rate is approximately 4.5 fps.

\begin{figure}[!t]
\centering
\includegraphics[width=\columnwidth]{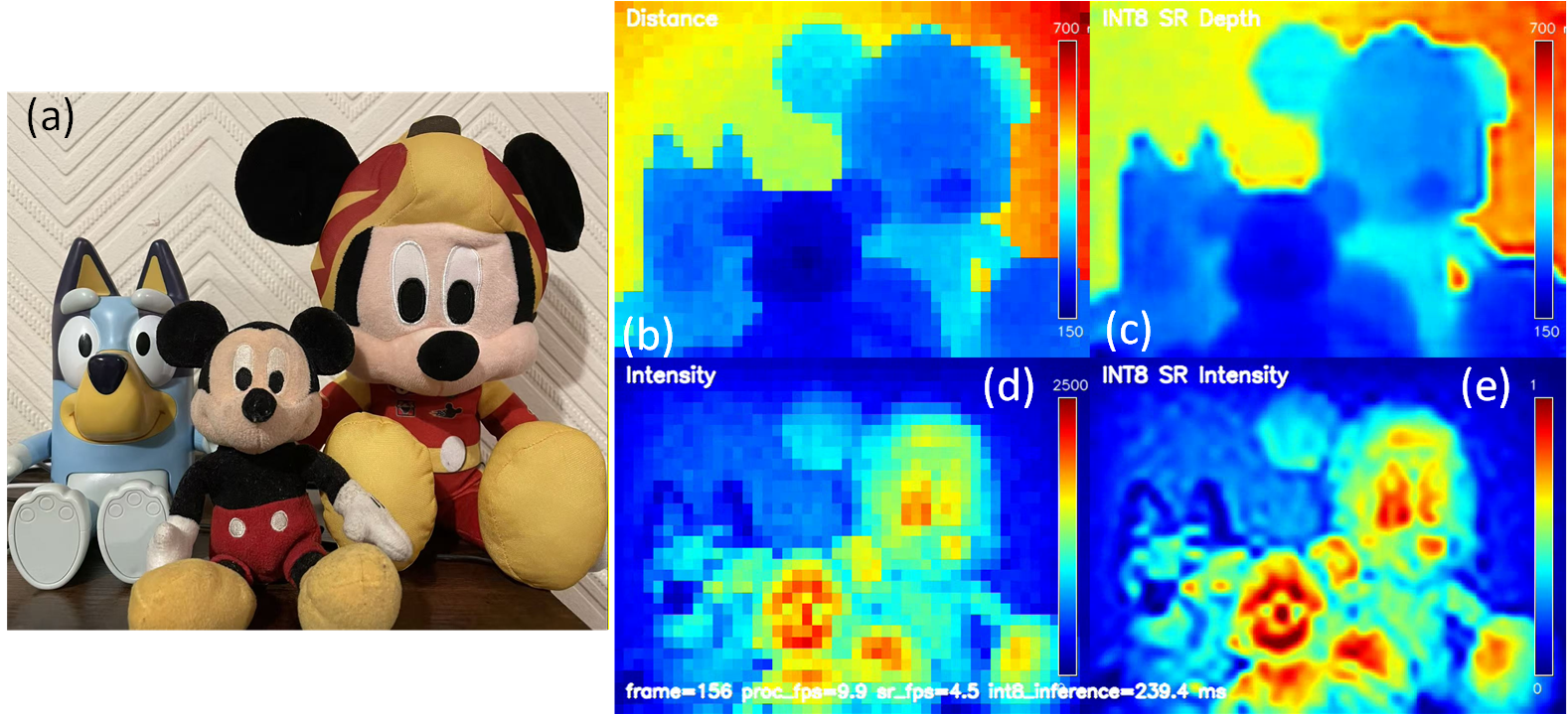}
\caption{Results on a stationary plush toy scene. (a) RGB reference photograph. (b) Raw LR depth map from the TMF8829 ToF sensor. (c) Upscaled depth map produced by LiteSR (INT8). (d) Raw LR intensity image. (e) Upscaled and normalized intensity image from LiteSR (INT8).}
\label{fig:stationary-plush-toy-results}
\end{figure}

The stationary scene test is shown in Fig.~\ref{fig:stationary-plush-toy-results}, which includes an RGB image and upscaled depth and intensity maps generated using the smallest LiteSR model, demonstrating improved spatial continuity and textural details. Live frame rate and per-frame inference latency are indicated. The corresponding video visualization is provided as Visualization~1, while Visualization~2 shows the real-time processing of a dynamic hand-puppet scene.

\begin{figure}[!t]
\centering
\includegraphics[width=\columnwidth]{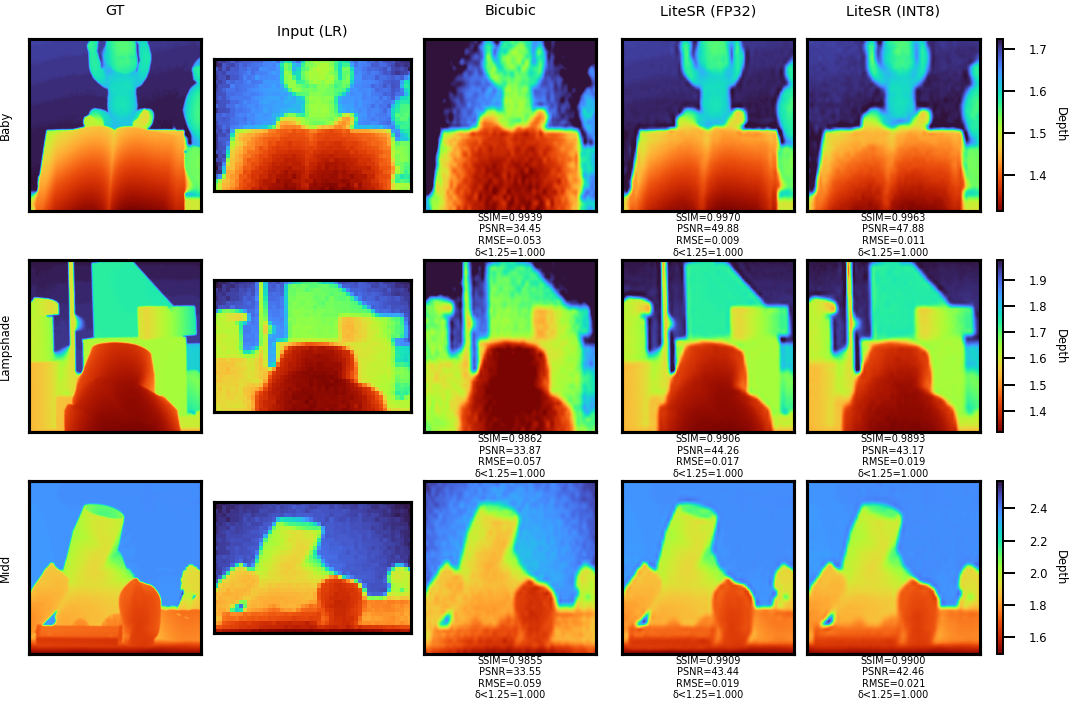}
\caption{Comparison of super-resolution results on depth test images for $8\times$ upscaling. From left to right: GT, LR, bicubic interpolation, LiteSR (FP32), and LiteSR (INT8).}
\label{fig:synthetic-depth-8x}
\end{figure}

\begin{table}[!t]
\caption{Model Complexity Comparison of LiteSR Under FP32 and INT8 Precision for $4\times$ and $8\times$ Upscaling}
\label{tab:model-complexity}
\centering
\begin{tabular}{|l|r|r|r|r|}
\hline
Model & Params & Param Size & MACs & OPs* \\
\hline
\multicolumn{5}{|c|}{Upscale ratio $4\times$ (256$\times$256)} \\
\hline
LiteSR FP32 & 355,949 & 1.36 MB & 2.043 G & 4.086 G \\
LiteSR INT8 & 93,133 & 363.80 KB & 0.973 G & 1.945 G \\
\hline
\multicolumn{5}{|c|}{Upscale ratio $8\times$ (512$\times$512)} \\
\hline
LiteSR FP32 & 402,221 & 1.53 MB & 5.148 G & 10.296 G \\
LiteSR INT8 & 139,405 & 544.55 KB & 4.077 G & 8.155 G \\
\hline
\end{tabular}
\begin{flushleft}
\footnotesize{*FLOPs for FP32, (integer operations per second) IOPs for INT8.}
\end{flushleft}
\end{table}

\begin{table*}[!t]
\caption{Ablation Study Evaluating the Contribution of EARB in the Depth Branch. Bold and underscored values indicate the better- and same-performing configurations for each metric. }
\label{tab:ablation-earb}
\centering
\begin{tabular}{|l|cc|cc|cc|cc|}
\hline
\multirow{2}{*}{Scene} & \multicolumn{2}{c|}{SSIM $\uparrow$} & \multicolumn{2}{c|}{PSNR (dB) $\uparrow$} & \multicolumn{2}{c|}{RMSE $\downarrow$} & \multicolumn{2}{c|}{$\delta < 1.25$ $\uparrow$} \\
\cline{2-9}
 & w/ & w/o & w/ & w/o & w/ & w/o & w/ & w/o \\
\hline
Art     & \textbf{0.930} & 0.929 & \textbf{36.26} & 36.17 & \textbf{0.043} & 0.044 & \underline{1.00} & \underline{1.00} \\
Books   & \underline{0.987} & \underline{0.987} & \textbf{45.12} & 44.89 & \underline{0.016} & \underline{0.016} & \underline{1.00} & \underline{1.00} \\
Bowling & \underline{0.982} & \underline{0.982} & \textbf{41.30} & 41.09 & \textbf{0.024} & 0.025 & \underline{1.00} & \underline{1.00} \\
Dolls   & \underline{0.988} & \underline{0.988} & \textbf{46.53} & 46.16 & \textbf{0.013} & 0.014 & \underline{1.00} & \underline{1.00} \\
Laundry & \textbf{0.979} & 0.978 & \textbf{42.22} & 41.91 & \underline{0.022} & \underline{0.022} & \underline{1.00} & \underline{1.00} \\
Moebius & \textbf{0.983} & 0.982 & \textbf{44.15} & 43.69 & \textbf{0.017} & 0.018 & \underline{1.00} & \underline{1.00} \\
Plastic & \underline{0.995} & \underline{0.995} & \textbf{49.32} & 48.86 & \underline{0.010} & \underline{0.010} & \underline{1.00} & \underline{1.00} \\
Reindeer& \underline{0.975} & \underline{0.975} & \textbf{40.51} & 40.39 & \textbf{0.026} & 0.027 & \underline{1.00} & \underline{1.00} \\
\hline
\end{tabular}
\end{table*}

\begin{table*}[!t]
\caption{Ablation Study Evaluating the Contribution of RLFB in the Intensity Branch. Bold and underscored values indicate the better- and same-performing configurations for each metric.}
\label{tab:ablation-rlfb}
\centering
\begin{tabular}{|l|cc|cc|cc|cc|}
\hline
\multirow{2}{*}{Scene} & \multicolumn{2}{c|}{MS-SSIM $\uparrow$} & \multicolumn{2}{c|}{PSNR (dB) $\uparrow$} & \multicolumn{2}{c|}{GMSD $\downarrow$} & \multicolumn{2}{c|}{nGMSD $\downarrow$} \\
\cline{2-9}
 & w/ & w/o & w/ & w/o & w/ & w/o & w/ & w/o \\
\hline
Art      & \textbf{0.910} & 0.906 & \textbf{24.47} & 24.34 & \underline{0.0014} & \underline{0.0014} & \textbf{0.0044} & 0.0046 \\
Books    & \textbf{0.912} & 0.910 & \textbf{23.55} & 23.43 & \underline{0.0019} & \underline{0.0019} & \textbf{0.0062} & 0.0064 \\
Bowling  & \textbf{0.963} & 0.961 & \textbf{30.42} & 30.20 & \underline{0.0006} & \underline{0.0006} & \textbf{0.0036} & 0.0038 \\
Dolls    & \textbf{0.906} & 0.902 & \textbf{23.47} & 23.28 & \textbf{0.0015} & 0.0016 & \textbf{0.0043} & 0.0045 \\
Laundry  & \textbf{0.867} & 0.864 & \textbf{21.72} & 21.64 & \textbf{0.0026} & 0.0027 & \textbf{0.0073} & 0.0074 \\
Moebius  & \textbf{0.914} & 0.910 & \textbf{25.92} & 25.81 & \underline{0.0008} & \underline{0.0008} & \textbf{0.0030} & 0.0031 \\
Plastic  & \textbf{0.961} & 0.958 & \textbf{26.98} & 26.71 & \textbf{0.0017} & 0.0018 & \textbf{0.0107} & 0.0114 \\
Reindeer & \textbf{0.942} & 0.937 & \textbf{26.11} & 25.74 & \textbf{0.0011} & 0.0012 & \textbf{0.0044} & 0.0047 \\
\hline
\end{tabular}
\end{table*}

\section{Discussion}
\subsection{Generalization to Higher Upscaling Ratios}
\begin{figure}[!t]
\centering
\includegraphics[width=\columnwidth]{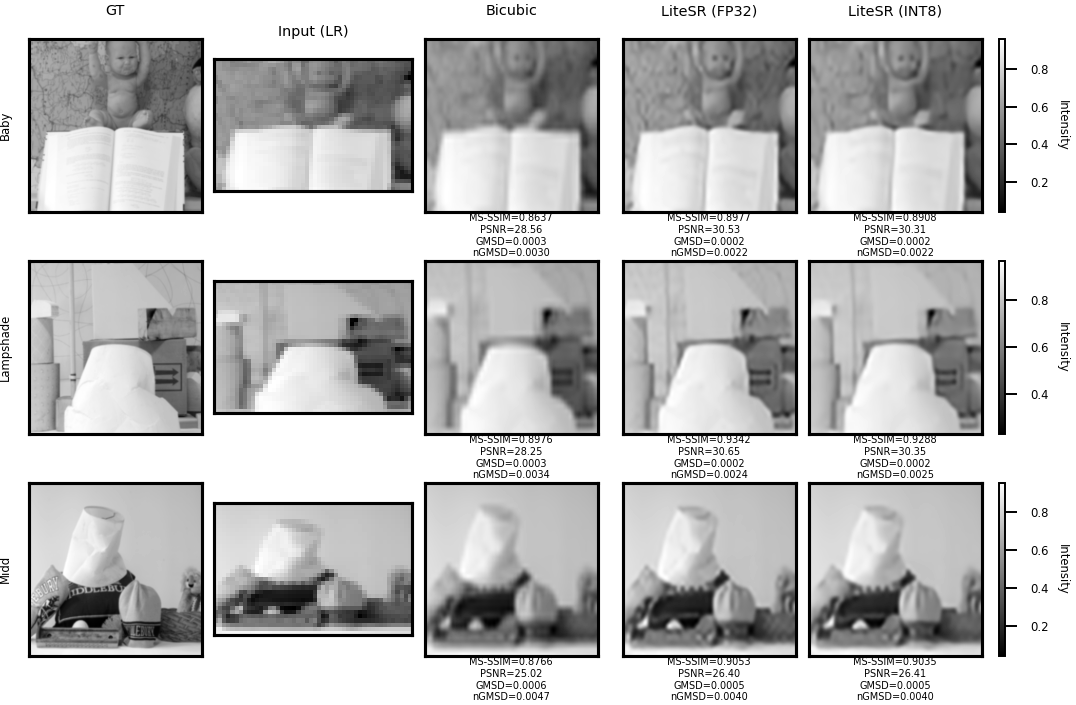}
\caption{Comparison of super-resolution results on intensity test images for $8\times$ upscaling. From left to right: GT, LR, bicubic interpolation, LiteSR (FP32), and LiteSR (INT8).}
\label{fig:synthetic-intensity-8x}
\end{figure}

To evaluate LiteSR under higher upscaling conditions, HR images of higher dimensions are generated for both training and testing. In addition to 256 $\times$ 256, an HR resolution of 512 $\times$ 512 is also considered. Reconstructed images from different scenes are presented in Figs.~\ref{fig:synthetic-intensity-8x} and \ref{fig:synthetic-depth-8x}. As the upscaling ratio increases from 4× to 8×, the performance gap between LiteSR and bicubic interpolation widens. LiteSR (FP32) achieves an average depth PSNR of 45.86 dB and MS-SSIM of 0.9124 for intensity at 8×8, compared to 33.96 dB and 0.8793 for bicubic interpolation. Bicubic interpolation exhibits noticeable blurring and loss of structural details at higher magnification factors, whereas LiteSR recovers sharper edges and more faithful intensity and depth structures, suggesting that as the upscaling ratio increases, the ill-posed nature of the SR problem intensifies, and the learned prior becomes increasingly critical for preserving structural fidelity.

The corresponding model (CH\_E=32, CH\_R=16, CH\_DH=10, and CH\_IH=8) shows that the INT8 quantized model achieves approximately a 74\% reduction in parameter storage and more than 50\% reduction in MACs compared to the FP32 baseline at 4× upscaling. At higher upscaling ratios, although the computational load increases, the INT8 model achieves substantial savings in both memory footprint and arithmetic operations. This reduces on-chip memory usage and off-chip memory access overhead, demonstrating that the quantization benefit scales consistently across upscaling ratios.

\subsection{Ablation Study}
The performance is evaluated by training with and without EARB modules for the depth branch, and RLFB modules for the intensity branch. The same test dataset is used across all ablation variants for a fair comparison. In each ablated variant, the removed module is replaced by a plain convolutional block with the same number of convolutional layers. The ablation results demonstrate that EARB consistently improves depth reconstruction across all eight test scenes, with modest but consistent gains. SSIM gains range from 0.001 to 0.004, and PSNR improvements of up to 0.46 dB are observed. The dilated convolution in EARB expands the receptive field without increasing the number of parameters, enabling the depth branch to capture larger spatial context at object boundaries. RLFB also consistently contributes to intensity reconstruction, with MS-SSIM improvements of up to 0.004 and PSNR gains of up to 0.37 dB. GMSD and nGMSD values are consistently lower with RLFB, confirming that the residual skip connection better preserves gradient structure than plain convolutions. The model sizes for the four ablation variants, i.e., the full model (EARB + RLFB), without EARB, without RLFB, and without both, are 1.36, 1.67, 1.40, and 1.72 MB, respectively.  Notably, the full model is the most compact among all variants, as EARB and RLFB employ 1×1 projection convolutions, which reduce the parameter count compared to the plain 3×3 convolutional replacements used in the ablated variants. This confirms that architectural efficiency and task performance are not mutually exclusive in the proposed design, making LiteSR well-suited for deployment on embedded systems.

\section{Conclusion}
This study presents LiteSR, a compact, fusion-free DL strategy for joint super-resolution of depth and intensity images acquired from a consumer-grade SPAD array, implemented on Arduino-class embedded hardware. Extensive evaluations on both synthetic and real datasets demonstrate that LiteSR achieves competitive SR quality while maintaining a compact model footprint suitable for edge deployment. The results suggest that DL-aided SR offers a practical and promising alternative to increasing pixel counts to achieve higher spatial resolution, a challenge that remains costly despite advances in semiconductor manufacturing and SPAD microarchitecture design. Leveraging the growing on-chip and on-board memory capacity of embedded hardware, together with modern DL compilation tools, DL-based SR can be deployed on edge devices without reliance on powerful GPUs. Future work will explore adaptive quantization and compression strategies that further reduce model size, and leverage the on-Arduino GPU and corresponding compilers to accelerate inference. Furthermore, histogram data are not exploited in this work; however, histograms are an important feature for capturing temporal photon information, and incorporating them could broaden the scope of applications. In particular, non-line-of-sight imaging and sensing through scattering media are potential directions enabled by histogram-based representations. Histogram-based SR in photon-starved environments would also constitute a valuable future research direction. Higher-resolution FLIM imaging could also be achieved using existing 8$\times$8 solutions, but with faster acquisition enabled by the larger array sizes of the 32$\times$48 and next-generation SPAD sensors.

\bibliographystyle{IEEEtran}
\bibliography{refs}

\end{document}